\documentstyle[11pt,jkas2]{article}
\input psfig.tex

\def\lsim{\lower.5ex\hbox{$\; \buildrel < \over \sim \;$}}
\def\gsim{\lower.5ex\hbox{$\; \buildrel > \over \sim \;$}} 
\begin{document}


\title{Spectral Properties of Galactic and Extragalactic Black Hole Candidates}
\author{Sandip K. Chakrabarti}
\address{Tata Institute of Fundamental Research, Mumbai, 400005  INDIA}

\abstract{We review current theoretical understanding of the
spectral properties (low and high states, transition of states, 
quasi-periodic  oscillations etc.) of the low mass as well
as supermassive black hole candidates.}


\maketitle  

Galactic black hole candidates show very interesting spectral properties:
the emitted power is sometimes in the softer energy 
(soft-state) and in other times it is in harder energy (hard state). 
In the low or the hard state, the soft bump is almost absent, and the energy
spectral index is $\alpha \sim 0.5-0.8$ in the range $2-50$kev 
($F(\nu) \sim \nu^{-\alpha}$). In the high or soft state the power-law
(hard) component is very weak, with a spectral slope of $\alpha \sim 1.5-1.8$.
(See, Ebisawa, Titarchuk \& Chakrabarti, 1996 for a list of black hole
candidates and their typical spectral indices.). 

Simplistic spherical symmetric models of Bondi 
flow of the 1950s (Bondi, 1952) or predominantly rotating flow
models of the 1970s (Shakura \& Sunyeav, 1973) or their other variations
cannot explain these strange properties. It
is recognized that the entire disk model has to be revised with
complete advection, heating and cooling effects so that one
could explain these spectra without invoking any external components
such as Compton cloud or corona etc. First such attempt was made by
Chakrabarti \& Titarchuk (1995, hereafter CT95; see, also Chakrabarti, 1996 
for a review) where concepts from the earlier solution on advective disks
(Chakrabarti, 1990; Chakrabarti \& Molteni, 1995) were used. Here,
it was pointed out that flows must deviate from a Keplerian disk
close to the horizon to satisfy inner boundary condition on the
horizon (radial velocity of matter equals to the velocity of
light and the corotating condition on the horizon). 
Disks on the equatorial plane may remain Keplerian until
very close to the black hole depending on accretion rate 
and viscosity parameter (the flow is bound with specific energy
${\cal E} <0$ everywhere) while away from the plane, 
the flow could have standing or oscillating shock 
waves (see, Chakrabarti et al., this volume). This is possible 
either because of lower viscosity, or because of weaker gravity
away from the plane, or because of energy deposition from the
Keplerian disk on the advecting corona, or in the wind-fed systems
where supersonic flow is deposited, or any combination of 
above. Thus the flow essentially has two components, one 
Keplerian and the other sub-Keplerian.  
\begin{figure}
\hbox{
\hskip 4.0cm
\vbox{
\vskip -5.0cm
\centerline{
\psfig{figure=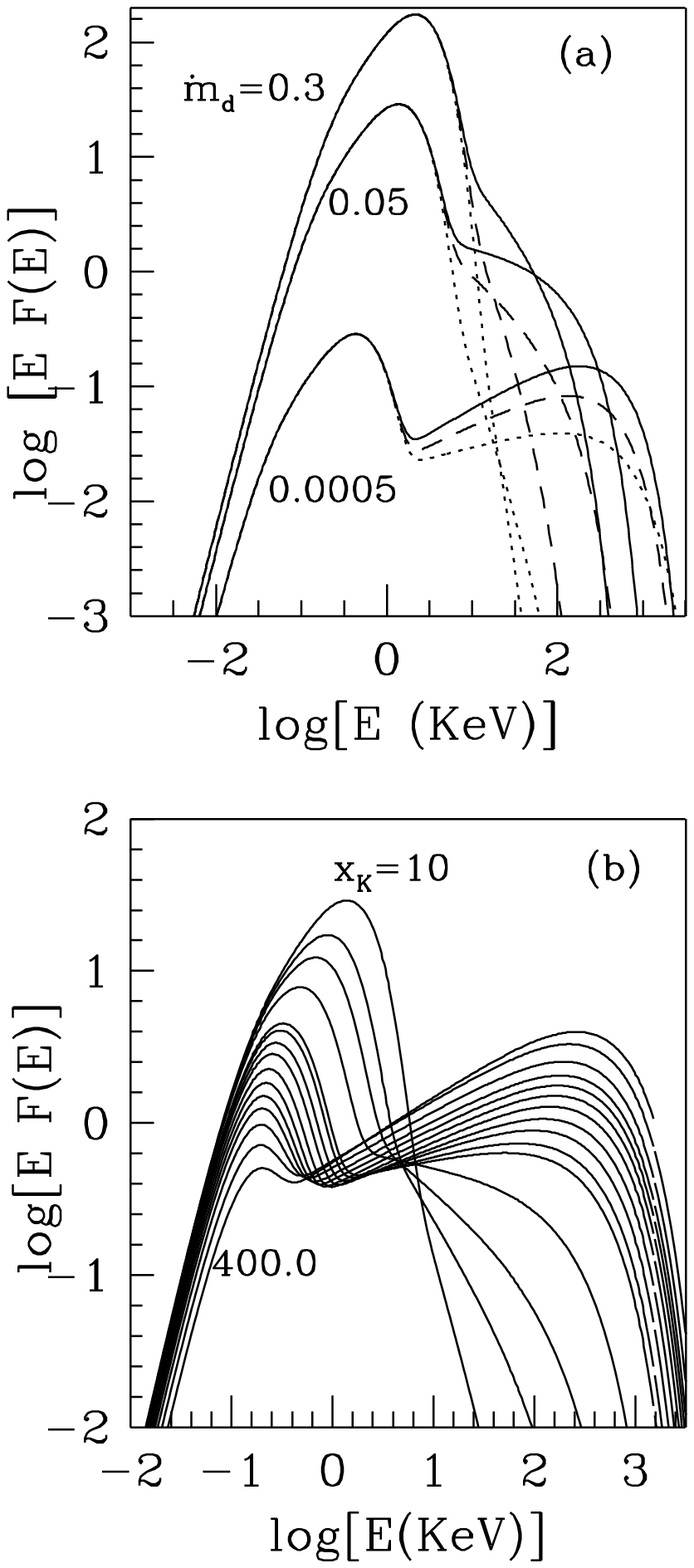,height=15.0 truecm,width=15 truecm,angle=0}}}}
\vspace {-1.0cm}
\noindent {\small {\bf Fig. 1(a-b)}: Model dependence of 
spectral properties. (a) Solid, long-dashed and 
short-dashed curves are for strong-shock, weak-shock and no-shock cases 
respectively. Strong shock cases produce harder spectra.
(b) Result of no-shock solution (${\dot m_{disk}}=0.05,\ {\dot m_{halo}}=1.0$). 
Spectrum changes from hard state to soft state as the Keplerian disk
approaches the black hole due to increase in $\alpha$ and ${\dot m}_d$ (as probably 
in the rising phase of a novae outburst).}
\end{figure}
This is required, since in many systems, the soft and the
hard components vary independently. If the Keplerian 
component rate is very small compared to the
sub-Keplerian rate, (typically, less than $0.1 {\dot M}_{Edd}$
for a sub-Keplerian rate of $1.0 {\dot M}_{Edd}$), the 
number of soft photons from the Keplerian disk
intercepted by the standing shocks is unable to cool 
the post-shock region and the object remains in a hard state. 
For higher Keplerian rate, the post-shock region cools dramatically 
and the object goes to the soft state. (Here, the phrase `post-shock'
really means the enhanced density region behind the centrifugal 
barrier experience by the inflow.) Even then, a weak hard tail is
produced not because of thermal Comptonization, but because of
bulk motion Comptonization as the relativistic matter rushes to 
the horizon and transfers bulk momentum to soft photons. Thus, the
weak hard tail is unique to black hole candidates and neutron star
candidates must not show this component. CT95
showed that for high enough rate (${\dot M} \gsim 3$) the
spectral slope is roughly $- 1.5$ while below that rate 
the power law is broken, since the hard component consists of
contributions from thermal and bulk motion Comptonizations.
\begin{figure}
\hbox{
\hskip 2.0cm
\vbox{
\vskip -5.0cm
\centerline{
\psfig{figure=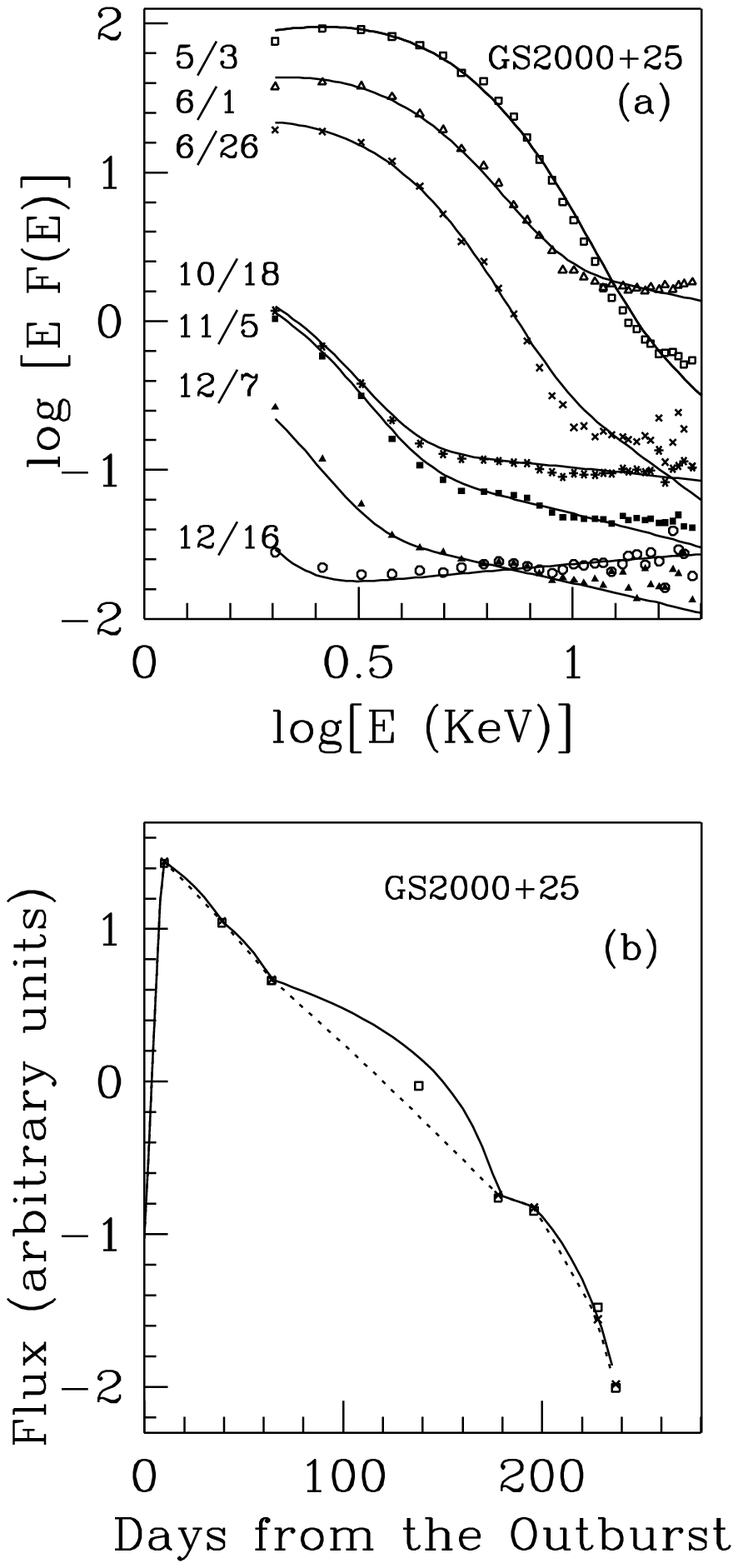,height=15.0 truecm,width=15.0 truecm,angle=0}}}}
\vspace {-1.0cm}
\noindent {\small {\bf Fig. 2(a-b)}: Rough fits of spectral evolution of X-ray novae
GS2000-25 using two component advective flow model and (b) comparison of derived 
light curve with the observed light curve. }
\end{figure}
Figs. 1(a-b) show results of different versions of CT95 solutions
of two component accretion.
In Fig. 1(a), the disk rate is changed while keeping the halo rate 
(${\dot m}_h={\dot m}_{Edd}$) fixed. The density enhancement
close to the black hole due to the centrifugal barrier may or may not
produce shock waves in the sub-Keplerian component. 
The solid, long-dashed and dotted curves are
for strong-shock, weak-shock and no-shock cases.
It is seen that the no-shock solutions mostly produce softer states
since the electron density is lower near the hole.
In Fig. 1b, a hard-state to soft-state transition is achieved
by changing the location where the disk component deviates from 
a Keplerian disk, while keeping the halo component fixed. This
behavior (achievable by increasing viscosity in the disk)
is expected to be important during the rising phase of 
a novae outburst.

Fig. 2a, drawn using data from Tanaka (1991), shows the spectral 
evolution of a typical X-ray novae GS2000+25 and its rough fit 
using two component advective disk model (TCAF) of Chakrabarti \& Titarchuk 
(1995) in the presence of a strong shock. Only 
parameters varied were the two mass accretion rates (for the disk and 
the halo components). The light curve (2-20keV) derived from these fitted
rates is shown in Fig. 2b. Linear-linear (solid) or linear-log (dotted)
interpolations of the rates were used. Squares and crosses are the
light curves from the observed data and fitted plots respectively.
Mass of the black hole (in Figs. 1ab and 2ab) was chosen to be $1M_\odot$ which 
after correction due to spectral hardening factor ($f\sim 1.9$,
see, Shimura \& Takahara, 1995) corresponds to a mass of $3.6M_\odot$.
The results remain similar when supermassive black holes are chosen.
The bumps after $\sim 60-70$d and after  $\sim 200$d of the outburst
are clearly reproduced. A single component model will have difficulty
to explain such variations as explain. Note that the light shows a decay
time scale of about $30$ days.

\begin{figure}
\vbox{
\vskip -5.0cm
\centerline{
\psfig{figure=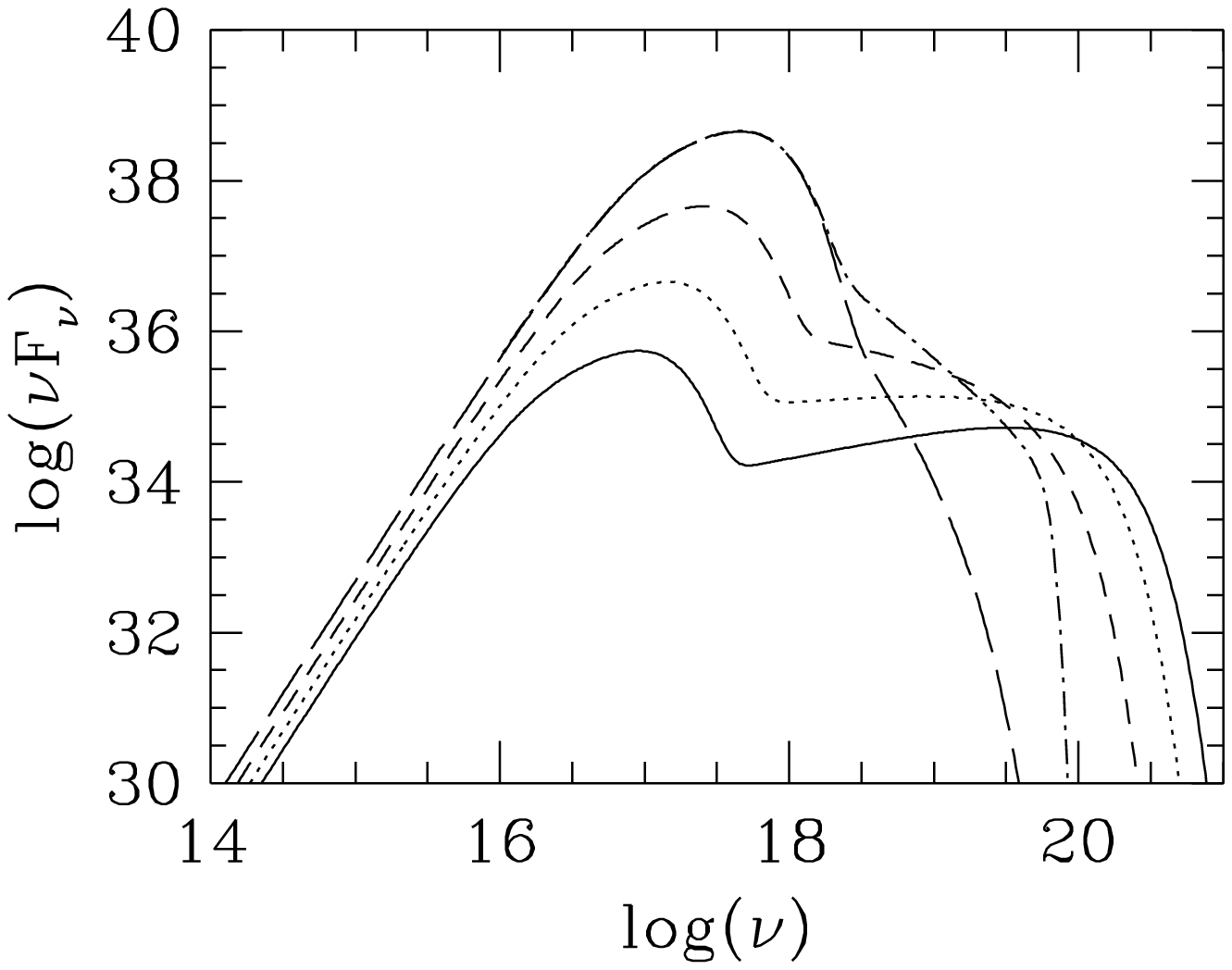,height=12.0 true cm,width=12.0 true cm,angle=0}}
\vskip -5.0cm
\centerline{
\psfig{figure=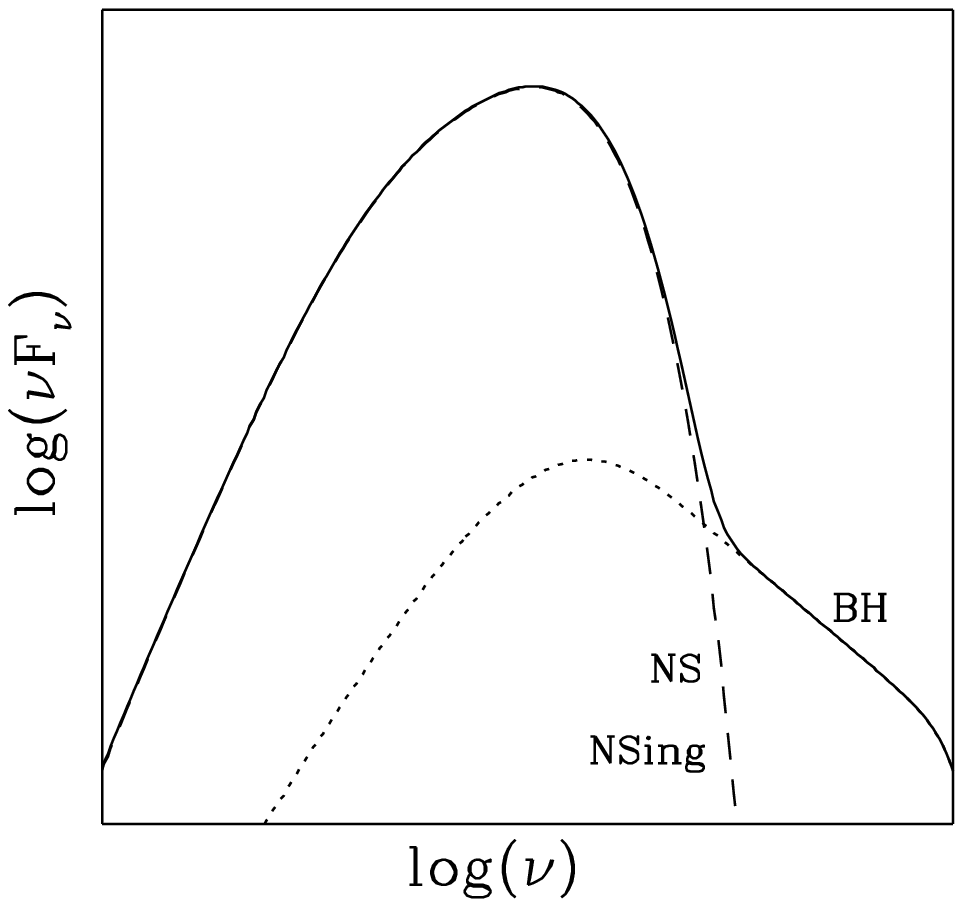,height=15.0truecm,width=15.0truecm,angle=0}}}
\vspace {-3.5cm}
\noindent {\small {\bf Fig. 3(a-b)}: 
Hard to soft state transition of a black hole
candidate of mass $5M_\odot$ as the Keplerian accretion rate is 
increased. (a) ${\dot m}_{disk}=0.001$ (solid), $0.01$ (long dashed), $0.1$
(short dashed) and $1$ (dotted) Dot-dashed curve is drawn after including
the bulk motion Comptonization. (b) Comparison of soft states of black hole
(BH), neutron stars (NS) and naked singularities (NSing). Black holes
always show weak hard tail even in the soft state, while other
compact bodies do not.}
\end{figure}

Figs. 3(a-b) shows CT95 strong shock solutions for variation
in the disk accretion rate ${\dot m}_{disk}$. The mass (uncorrected for spectral hardening)
of the black hole and the shock location were chosen to be $5M_\odot$,
and $x_s=10 x_g$ (typical values) respectively.
For higher mass (QSOs and AGNs) black holes, the result is
similar, though much cooler (electron temperature $T_e \propto M^{0.04-0.1}$).
$\nu F(\nu)$ is plotted against $\nu$. In Fig. 3a, the
object goes from hard to soft states, as the Keplerian accretion rate
is increased from $0.001$ to $1.0$ while keeping the sub-Keplerian
rate fixed at $1 {M_{Edd}}$. In Fig. 3b, we show the spectral shape in
soft states. The weak hard tail of slope $\alpha \sim -1.5$
is formed due to relativistic radial motion of the flow outside the 
horizon of a black hole (marked as BH). The neutron stars (NS) 
or naked singularities (NSing) should not have this hard tail (CT95).

\begin{figure}
\vbox{
\centerline{
\psfig{figure=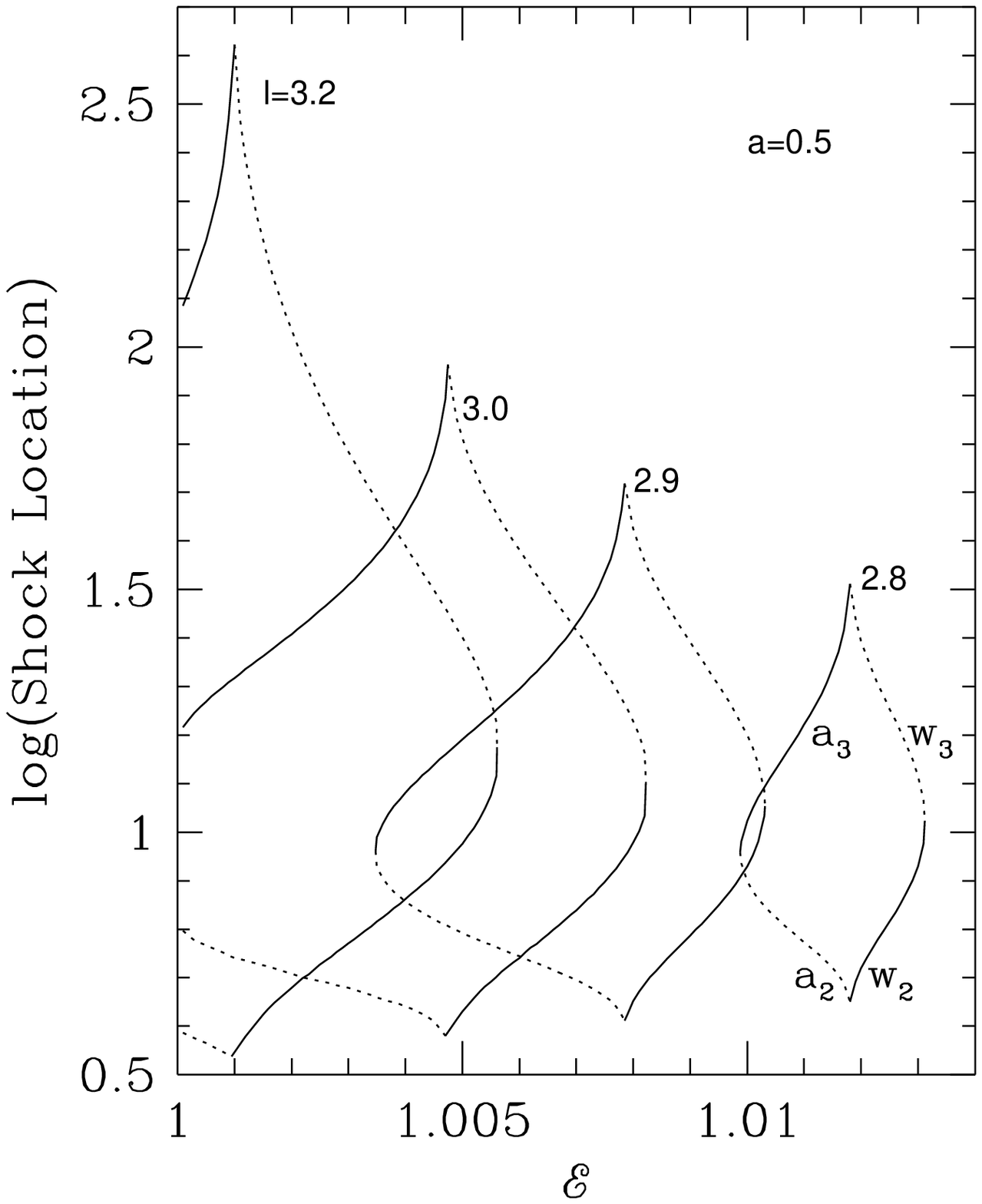,height=11 truecm,width=11 truecm,angle=0}}}
\vspace {-2cm}
\centerline{
\psfig{figure=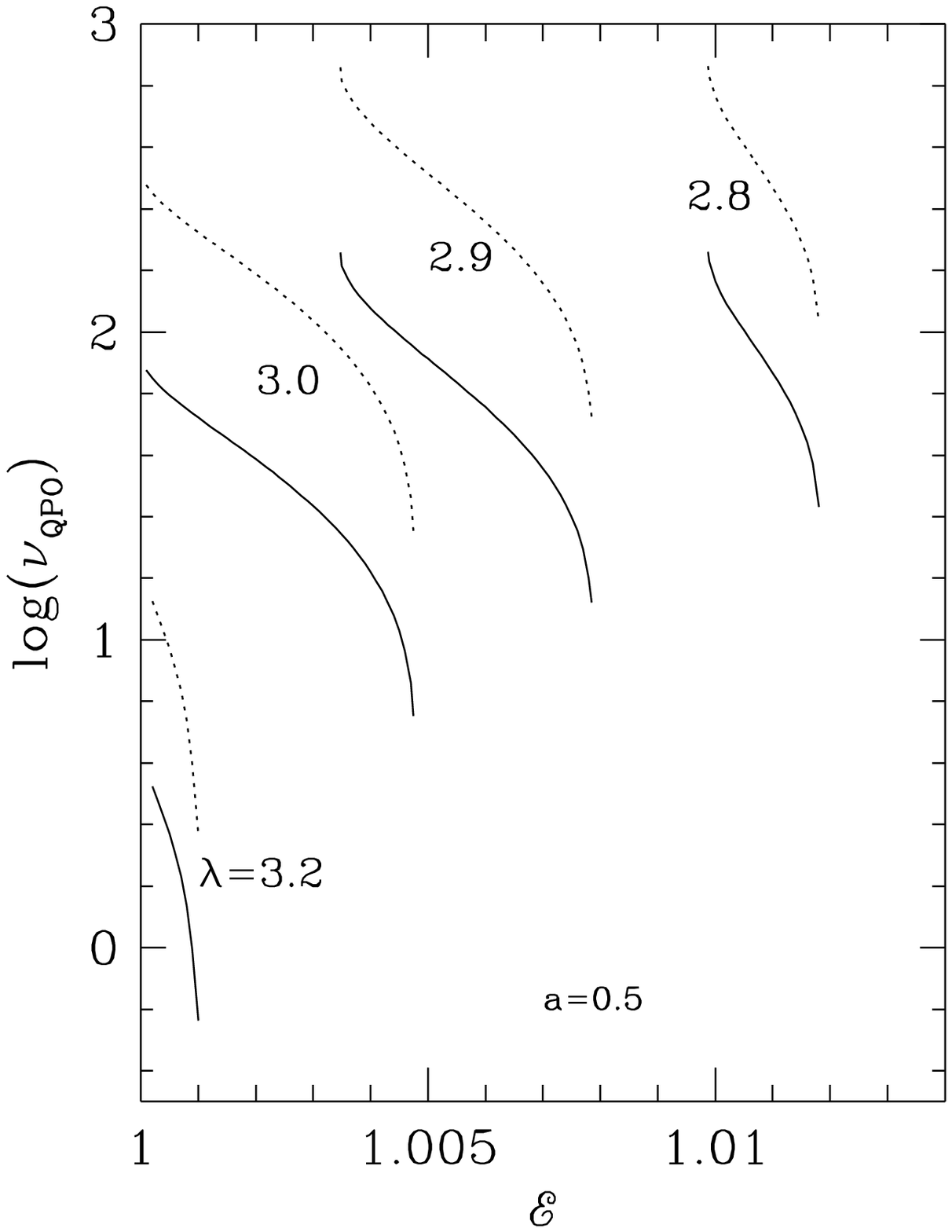,height=10truecm,width=10truecm,angle=0}}
\vspace {-1.0cm}
{\small {\bf Fig. 4(a-b)}: Variation of (a) shock locations (solid
curves $a_3$ and $w_2$ are stable in accretion and winds)
and (b) quasi-periodic oscillation frequencies as functions of the
angular momentum and energy of the flow. Kerr parameter $a=0.5$ is
chosen. Solid curves in (b) are drawn for $a_3$ and dotted curves are
drawn assuming that the shocks are absent but Keplerian disk joins with
sub-Keplerian flow at $a_3$.}
\end{figure}

Many of the black hole candidates also show the so-called quasi-periodic
oscillations in the range of several millihertz to some kilohertz range
(e.g., van der Hooft et al. 1996; Grebenev et al., 1993)
with the hard X-rays modulated by 10-30\% or more.
As shown by Ryu et al. (1996) and Molteni et al (1996), these
oscillations could be well understood by the resonance 
oscillation or the dynamic oscillation of the high density
region near the centrifugal barrier (see Chakrabarti
et al. this volume). Typically, the frequency of oscillation $f\sim 
4 x_s^{3/2} (2GM/c^3)$ s. In Fig. 4a, we show the shock location
(Kerr black hole parameter $a=0.5$) for angular momentum of 
matter $\lambda=3.2$, $3.0$, $2.9$, $2.8$ respectively
(Chakrabarti, 1996b). In Fig. 4b,
we show the corresponding QPO frequencies. Solid lines are for shock 
oscillations and dashed lines are for oscillation of the same flow
if shocks were absent (shock free solutions) and the Keplerian
disk joined the sub-Keplerian  at $x_{tr}=x_s$ instead.

\noindent To appear in the Proceedings of 6th Asia-Pacific Conference
(Journal of Korean Astronomical Society) August, 1996, Eds. H.M. Lee and S. Kim

\noindent Authors's address AFTER November 26th, 1996:\\

\noindent Prof. S.K. Chakrabarti\\
\noindent S.N. Bose National Center for Basic Sciences\\
\noindent JD Block, Sector -III, Salt Lake\\
\noindent Calcutta 700091, INDIA\\

\noindent e-mail: chakraba@bose.ernet.in  OR chakraba@tifrc2.tifr.res.in \\

\end{document}